\documentclass[fleqn,usenatbib]{mnras}

\usepackage{newtxtext,newtxmath}
\usepackage[T1]{fontenc}
\usepackage{ae,aecompl}

\usepackage{graphicx}	
\usepackage{amsmath}	
\usepackage{amssymb}	

\defcitealias{Zuc2015}{Paper~I}
\defcitealias{Zuc2016}{Paper~II}

\title[Phase Distance Correlation Periodogram]{Detection of Periodicity Based on Independence Tests -- III. Phase Distance Correlation Periodogram}

\author[S. Zucker]{
Shay Zucker$^{1}$\thanks{E-mail: shayz@post.tau.ac.il} \\
$^{1}$School of Geosciences, Raymond and Beverly Sackler Faculty of 
Exact Sciences, Tel Aviv University, Tel Aviv, 6997801, Israel}

\date{Accepted XXX. Received YYY; in original form ZZZ}

\pubyear{2017}

\begin{document}
\label{firstpage}
\pagerange{\pageref{firstpage}--\pageref{lastpage}}
\maketitle

\begin{abstract}
I present the Phase Distance Correlation (PDC) periodogram -- a new periodicity metric, based on the Distance Correlation concept of G\'{a}bor Sz\'{e}kely. For each trial period PDC calculates the distance correlation between the data samples and their phases. PDC requires adaptation of the Sz\'{e}kely's distance correlation to circular variables (phases). The resulting periodicity metric is best suited to sparse datasets, and it performs better than other methods for sawtooth-like periodicities. These include Cepheid and RR-Lyrae light curves, as well as radial velocity curves of eccentric spectroscopic binaries. The performance of the PDC periodogram in other contexts is almost as good as that of the Generalized Lomb-Scargle periodogram. The concept of phase distance correlation can be adapted also to astrometric data, and it has the potential to be suitable also for large evenly-spaced datasets, after some algorithmic perfection.
\end{abstract}




\section{Introduction}
\label{sec:intro}

In the two previous papers of this series (\citealt{Zuc2015}, hereafter \citetalias{Zuc2015}; \citealt{Zuc2016}, hereafter \citetalias{Zuc2016}) I have introduced a new non-parametric approach to the detection of periodicities in sparse datasets.  The new approach I introduced in those two papers followed the logic of string-length techniques \citep[e.g.,][]{LafKin1965,Cla2002} and quantified the dependence between consecutive phase-folded samples  (serial dependence), for every trial period. While the classic string-length methods essentially quantified the linear (Pearson) correlation in those pairs of samples, the new approach I introduced measured the generalized dependence thereof. The improved technique I introduced in Paper II used the Blum-Kiefer-Rosenblatt (BKR) statistic \citep*{Bluetal1961} to quantify this dependence, constituting an improvement over the use of the Hoeffding statistic \citep{Hoe1948} in Paper I. The BKR approach proved to be particularly successful for periodicities of the sawtooth type, including also radial-velocity (RV) curves of eccentric single-lined spectroscopic binaries (SB1).

The most basic and widely used technique to search for periodicities in unevenly-sampled data is the Lomb-Scargle periodogram \citep{Lom1976,Sca1982}. The Lomb-Scargle periodogram is explicitly based on searching for sinusoidal periodicities, usually assuming that every periodicity can be represented as a series of sinusoids (harmonics). Lomb-Scargle periodogram has been improved and generalized in several ways \citep[e.g.,][]{Bre2001a,Bre2001b}, and \citet{Van2017} presents an excellent and intuitive introduction to this staple of astronomical data analysis. Here I will refer by the name Generalized Lomb-Scargle (GLS) periodogram to the method introduced by \citet{ZecKur2009} which simply considers a constant offset term in addition to the sinusoid.

The inspiration to the current work is the observation that the GLS can be represented also as a correlation measure. For a specific trial period, the GLS value is essentially the correlation between the samples and their corresponding phases according to the trial period. Since the phase is a circular variable \citep[e.g.][]{MarJup2000}, the correlation is not calculated in the usual way applicable to two linear variables. \citet{Mar1976} suggested a statistic to measure the correlation between a linear variable and a circular one. A simple algebraic manipulation shows that the statistic 
\citeauthor{Mar1976} introduced leads exactly to the GLS formula.

\citeauthor{Mar1976}'s statistic measures a linear-circular correlation. The basic change I propose here is to use a measure of a linear-circular {\it dependence}, rather than {\it correlation}. The way forward is provided by a recent progress in the estimation of dependence by \citet*{Szeetal2007}. \citeauthor{Szeetal2007} developed a measure of dependence between two variables (not necessarily of the same dimension), which they dubbed {\it distance correlation}. The distance correlation is zero if and only if the two variables are statistically independent, unlike the usual Pearson correlation statistic, which can be zero for dependent variables. In its usual formulation, the calculation of the distance correlation has an appealing resemblance to Pearson correlation, and can therefore be considered quite intuitive and elegant. 

The only missing step is the adaptation of the distance correlation to circular variables.  I have applied the derivation of \citet{Szeetal2007} assuming one of the two variables is circular, and arrived at a satisfactory prescription. Thus, calculating the distance correlation between the samples and their phases at each trial period, amounts to a generalization of the GLS to general dependence -- the Phase Distance Correlation (PDC) Periodogram.

While in Section~\ref{sec:PDC} I present the final expression I obtained, I justify it a little more rigorously in the Appendix. In Section~\ref{sec:comp} I present a comparison of  the performance of the new periodogram against some other alternatives. Finally I discuss those results and future work in Section~\ref{sec:disc}.

\section{Phase Distance Correlation}
\label{sec:PDC}

In what follows I detail the recipe to calculate the phase distance correlation for a given trial period. Most of the calculation is identical to the calculation of the distance correlation in \citet{Szeetal2007}, except for the definition of the phase distance, which is an adaptation of the linear distance to a circular variable. In the appendix I justify this expression based on a rigorous derivation.

Let  us assume our dataset consists of $N$ samples $x_i$ ($i=1,...,N$), taken at times $t_i$, and we wish to calculate the phase distance correlation for the trial period $P$.

First, let us compute an $N$ by $N$ sample distance matrix:
\begin{equation}
a_{ij} = |x_i-x_j|\,.
\end{equation}

In order to compute a phase distance matrix we start by calculating a phase difference matrix:
\begin{equation}
\phi_{ij} = (t_i-t_j) \mod{P}
\end{equation}
which we then convert to a phase distance matrix:
\begin{equation}
\label{eq:phd}
b_{ij} = \phi_{ij}(P-\phi_{ij})\,.
\end{equation}

This choice of expression for phase distance is not trivial. We might have thought about other, maybe more intuitive expressions, such as $\min(\phi_{ij},P-\phi_{ij})$, or $\sin({\pi\phi_{ij}/P})$ \citep[e.g.][]{MarJup2000}. However, this expression is the result of applying the derivation of \citeauthor{Szeetal2007} to circular variables. In fact, I experimented with several other phase distance alternatives, and none yielded better performance than the one in Equation~\ref{eq:phd}.

The next step in the calculation is 'double centering' of the matrices $a$ and $b$, i.e.,
\begin{equation}
\label{eq:distances}
A_{ij} = a_{ij}-a_{i\cdot}-a_{\cdot j}+a_{\cdot\cdot} \\
B_{ij} = b_{ij}-b_{i\cdot}-b_{\cdot j}+b_{\cdot\cdot}
\end{equation}
where $a_{i\cdot}$ is the $i$-th row mean, $a_{\cdot j}$ is the $j$-th column mean, and $a_{\cdot\cdot}$ is the grand mean of the $a$ matrix. Similar definitions apply for the $b$ matrix. $A_{ij}$ and $B_{ij}$ are now the centered distance matrices. We can now finally define the dependence measure by:
\begin{equation}
\label{eq:cor}
\mathrm{PDC} = \frac{\sum\limits_{ij}A_{ij}B_{ij}}{\sqrt{(\sum\limits_{ij}A^2_{ij})(\sum\limits_{ij}B^2_{ij})}}
\end{equation}
Superficially, if we ignore the dependencies among the entries of the matrices $A$ and $B$, the numerator in Eq.~\ref{eq:cor} resembles the covariance between $A_{ij}$ and $B_{ij}$, and the whole expression resembles the Pearson correlation. This resemblance led \citeauthor{Szeetal2007} to choose the names distance covariance and distance correlation.

$\mathrm{PDC}$ is a dimensionless quantity, and it is bounded below by $0$ (complete phase independence), and above by $1$. Higher values mean stronger dependence on phase, and therefore indicate a periodicity is more likely.

I chose to show here two examples of applying the PDC periodogram to simulated data. In both examples I drew random times between $0$ and $1000$ days, from a uniform distribution, and generated a periodic signal with a period of two days. In both examples I calculated the PDC and the GLS for comparison, for a uniform frequency grid ranging between $10^{-4}$ and $1\,\mathrm{day}^{-1}$, with a constant spacing of $10^{-4}\,\mathrm{day}^{-1}$.

\begin{figure}
	\includegraphics[width=\columnwidth]{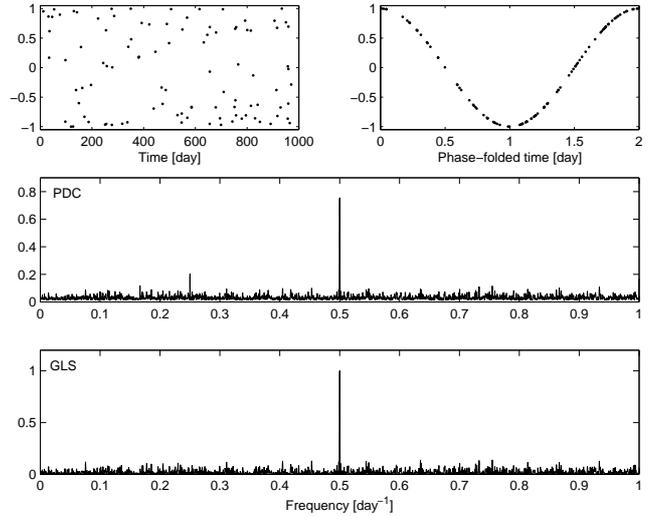}
    \caption{An example application of the PDC periodogram to a $100$-sample dataset, containing a pure sinusoid. The two upper panels show the data -- raw (left panel) and phase folded (right panel). The middle panel shows the PDC and the lower panel shows the GLS.}
    \label{fig:example1}
\end{figure}

Figure~\ref{fig:example1} presents a case of $100$ samples, and a pure sinusoidal signal, with no added noise. This is an easy case to detect, especially with the GLS, which is actually based on the assumption of sinusoidal signal. As expected, both periodograms manage to detect the signal very easily -- a very prominent and sharp peak at the correct frequency is apparent. Moreover, it seems that the two periodograms are very similar, except for a small peak corresponding to a subharmonic of the true frequency, which appears in the PDC plot.

\begin{figure}
	\includegraphics[width=\columnwidth]{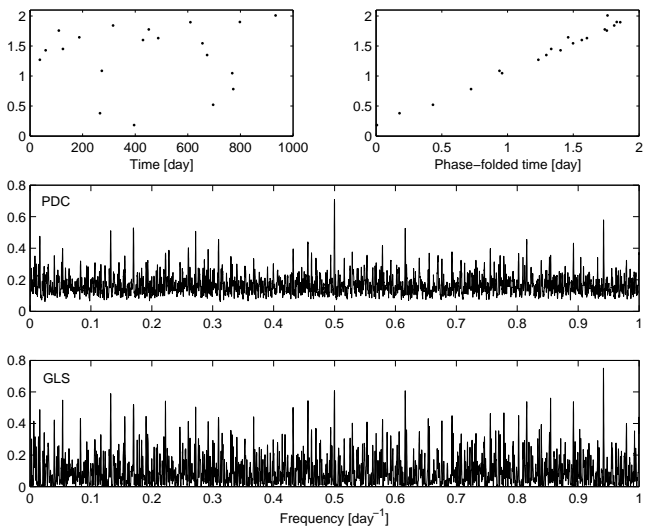}
    \caption{An example application of the PDC periodogram to a $20$-sample dataset, containing a sawtooth signal, with added Gaussian noise with an SNR of $2$. The two upper panels show the data -- raw (left panel) and phase folded (right panel). The middle panel shows the PDC and the lower panel shows the GLS.}
    \label{fig:example2}
\end{figure}

Figure~\ref{fig:example2} presents a much more difficult case, with only $20$ samples, a sawtooth signal, and added noise, at a signal-to-noise ratio (SNR) of $2$  (See \citetalias{Zuc2015} for an exact definition of SNR in the current context). the PDC plot exhibits a decisive peak at the correct frequency, while no such prominent peaks appear in the GLS plot. The highest peak in the GLS periodogram actually appears at a completely wrong frequency. In the next section I will try to quantify how typical this case is.

\section{Comparison}
\label{sec:comp}

I present here a performance comparison of PDC against three other periodicity metrics. I already mentioned the first one -- GLS -- in the previous section. The GLS is obviously a kind of gold standard in the field of periodicity searches in unevenly-sampled datasets, against which every new method should be compared. Next is the serial von-Neumann Ratio \citep{vonetal1941}, which I also included in the tests I performed in \citetalias{Zuc2015} and \citetalias{Zuc2016}. I use the periodicity metric based on the von-Neumann Ratio as a representative of the various string-length techniques. The third metric I use for benchmarking is the BKR-based approach I introduced in \citetalias{Zuc2016}.

The performance diagnostic I use is the Receiver Operating Characteristic (ROC) curve \citep[e.g.][]{Faw2006}. The ROC curve of a detection scheme is simply a plot of the True Positive Rate (or sensitivity), against the False Positive Rate (or type I error). As we change the detection threshold we change both of those two rates, and the dependence between them is monotonic. 

In all the tests I ran I used the same procedure to draw the observation times as I did in the examples I have shown above -- uniform distribution between $0$ and $1000\,\mathrm{days}$. I also calculated the periodicity metrics on the same frequency grid with $10^4$ frequencies between $10^{-4}$ and $1\,\mathrm{day}^{-1}$. 

For each number of samples I wished to test I first drew $1000$ realizations of light curves with random times and uncorrelated Gaussian random samples. I then calculated the four periodograms for each such realization of white noise, and normalized them (i.e., subtracted the means and divided by the standard deviations) to get Z~scores. For each periodogram I got an empirical distribution of the maximum Z-score values. The percentiles of that distribution served as thresholds for a given FPR.

I could now use those thresholds to test the periodograms on light curves that included periodic signals. I used six kinds of periodic signals -- the same ones I used in \citetalias{Zuc2015} and \citetalias{Zuc2016} -- pure sinusoid, almost sinusoidal, sawtooth, box shape ('pulse'), eccentric eclipsing binary (EB) and eccentric SB1 RV curve. I simulated always the same period -- two days. For a given configuration of periodicity shape, number of samples, and SNR, I tested $1000$ realizations to estimate the true positive rate (TPR). I standardized each periodogram to get the corresponding Z~scores, and applied the thresholds I had previously obtained from the uncorrelated Gaussian random samples. Thus each threshold yielded a pair of FPR and TPR values, resulting in the ROC curve.

\begin{figure}
	\includegraphics[width=\columnwidth]{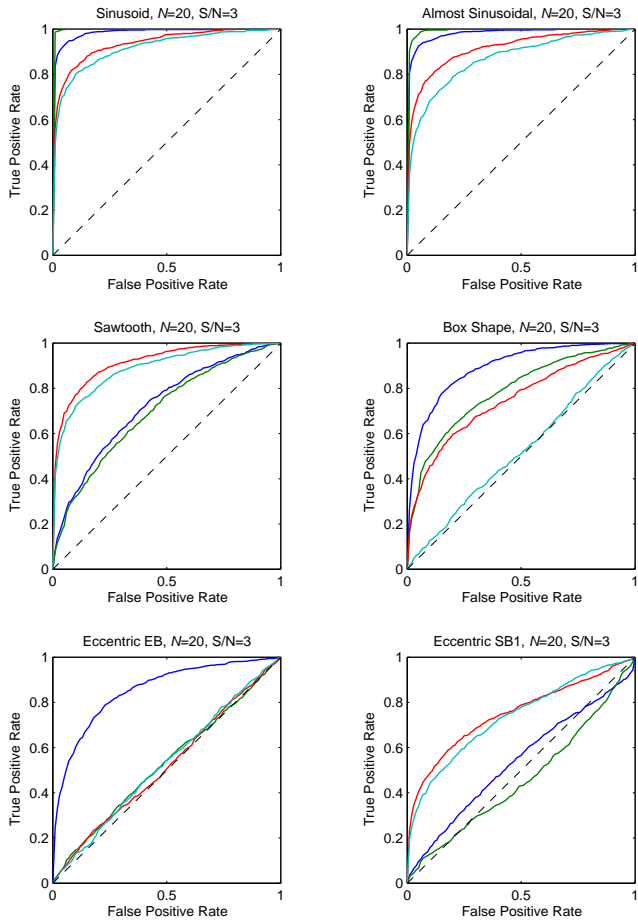}
    \caption{ROC curves for the four periodicity metrics and six periodicity shapes, for light curves with $20$ samples and SNR of $3$. Red lines represent the PDC, green lines represent the GLS, dark blue stands for the von-Neumann ratio and light blue represents the BKR.}
    \label{fig:all_3_20}
\end{figure}

A close look at Fig.~1 of \citetalias{Zuc2016} reveals that the situation which differentiates the most between the various methods is having a $20$-sample dataset, with an SNR of $3$. Therefore I present in Figure~\ref{fig:all_3_20} the ROC curves obtained for datasets with $20$ samples and SNR of $3$. The ROC curves of PDC are represented by red lines, those of the GLS by green lines, the von-Neumann ratio by dark blue, and the BKR periodicity metric by light blue.

In the case of pure sinusoid (upper left panel), as expected, the GLS dominates over all other methods. PDC performs only slightly better than the BKR periodogram, which has the poorest performance. This is no surprise, as the GLS is indeed expected to have the best performance for sinusoidal signals. The situation with the 'almost sinusoidal' periodicity shape is not much different. 

PDC turns out to be the best performer for sawtooth-like signals, better than BKR, which was previously shown to have the best performance for those signals in \citetalias{Zuc2016}. BKR performs poorly for box-shape periodicity, as I have already shown in \citetalias{Zuc2016}, but PDC is much better than BKR, and almost approaches the performance of the GLS in that situation. For eccentric EB, there seems not to be a significant difference between the failure of most methods, except for the von-Neumann periodogram which performs much better. For an RV curve of eccentric SB1, PDC performs almost like the BKR, except for a small advantage for PDC in the low FPR side.

\section{Discussion}
\label{sec:disc}

I introduced here the PDC periodogram -- a periodogram which generalizes the GLS to shapes other than sinusoidal.PDC is essentially a generalization of the GLS correlation approach to general dependence, and it has an appealing algebraic structure similar to correlation. 

In all situations I presented in Figure~\ref{fig:all_3_20}, the PDC periodogram outperformed the BKR periodogram, which I presented in \citetalias{Zuc2016}, and thus constitutes a clear improvement. Unlike BKR, whose main advantage is for sawtooth-like periodicities, PDC also performs almost as well as the GLS in other circumstances. Sawtooth-like periodicities are typical in photometry to Cepheids and RR-Lyrae stars, and in spectroscopy to RV curves of eccentric SB1 and exoplanets. Recently, \citet{Pinetal2017} highlighted the importance of improving the capability to detect periodicities in RV signals, especially for eccentric orbits of binary stars and exoplanets. The PDC periodogram serves to achieve this goal.

The original distance correlation which \citet{Szeetal2007} presented, is well defined also for multidimensional variables. Thus, PDC, which is based on the distance correlation, can be extended also to two-dimensional signals. Specifically, it can be extended to astrometric signals. Currently, the astrometric detection of Keplerian orbits (of exoplanets and binary stars) is based on a somewhat na\"{i}ve extension of the GLS to two-dimensional signals \citep*[\citealt{Sozetal2003};][]{Baretal2009}. This approach performs poorly for eccentric orbits. A natural extension of the PDC periodogram to two-dimensional signals (using Euclidean distance for the two-dimensional variable) will probably provide a much more natural solution.

It will be interesting to examine also the performance of the PDC periodogram under non-random uneven sampling laws, such as those affected by periodic observability constraints, or when the signal comprises two different periodicities, e.g. an RV curve of a two-planet system.

The main drawback of the recipe I presented in Section~\ref{sec:PDC} is its heavy computational burden. It involves $O(N^2)$ calculation for each trial period. This currently limits the application of PDC to sparse datasets. However, sparse data is the relevant context for the PDC periodogram, since in large datasets periodicities are usually adequately detectable by the GLS or other conventional techniques. Nevertheless, one approach to try and improve the complexity of the computation is suggested in \citet{HuoSze2016}. \citeauthor{HuoSze2016} use an unbiased version of distance correlation first suggested by \citet{SzeRiz2014}, and apply to it an AVL-tree computational approach \citep{AdeVel1962}, to obtain an $O(N\log N)$ algorithm to calculate the distance correlation. Future work should include an attempt to apply this algorithm to PDC and increase the allowed size of the dataset.

Another promising research direction concerns evenly-spaced time series, such as {\it Kepler} light curves \citep{Boretal2010}. In this case the phase distance matrix $b_{ij}$ becomes a Toeplitz matrix and may be amenable to significant computational shortcuts \citep[e.g.][]{Vic2001}, especially for periods that are integer multiples of the sampling interval. Thus, specific periods may result in even more degenerate matrix structures, potentially constituting a set of 'natural' periods. This may pave way to a Fourier-like transform, one that is based on general periodicities, and not focusing on sinusoidal ones.

In summary, the phase distance correlation approach I present here, has a potential to contribute significantly to solve the challenge of finding general periodicities, both in unevenly-sampled sparse datasets and in large evenly-sampled ones.

\section*{Acknowledgements}
This research was supported by the ISRAEL SCIENCE FOUNDATION (grant No. 848/16).

\appendix
\section{Phase Distance Justification}
\label{app}

\citet{Szeetal2007} based their derivation of the distance correlation on the probability-theory concept of the {\it characteristic function}. The characteristic function $\varphi_X$ of a linear random variable $X$ is essentially equivalent to a Fourier transform of its probability density function $f_X$:
\begin{equation}
\varphi_X(t) = \int_\mathbb{R} e^{itx} f_X(x) \mathrm{d}x
\end{equation}
The joint characteristic function $\varphi_{XY}(t,s)$ of the two random variables is similarly obtained from the joint probability density function $f_{XY}(x,y)$. Statistical independence of the two variables mean $f_{XY}(x,y)=f_X(x)f_Y(y)$, which is known to be equivalent to $\varphi_{XY}(t,s)=\varphi_X(t)\varphi_Y(s)$.

Essentially, \citeauthor{Szeetal2007} proposed the following dependence measure between the two linear random variables $X$ and $Y$:
\begin{equation}
\mathcal{V}^2(X,Y) = \frac{1}{\pi^2} \int_{\mathbb{R}^2} \frac{|\varphi_{XY}(t,s) - \varphi_X(t) \varphi_Y(s)|^2}{t^2s^2} \, \mathrm{d}t \, \mathrm{d}s
\end{equation}
\citeauthor{Szeetal2007} dub this measure distance covariance. Originally, they considered multidimensional variables, but for our needs here we focus only on one-dimensional variables.

Given a finite bivariate sample: $\{(x_k,y_k)\}_{k=1}^N$, the empirical characteristic function of $X$ is defined by:
\begin{equation}
\label{eq:ecfx}
\varphi^N_X(t) = \frac{1}{N} \sum_{k=1}^N e^{itx_k}
\end{equation}
and the bivariate joint empirical characteristic function is:
\begin{equation}
\varphi^N_{XY}(t,s) = \frac{1}{N} \sum_{k=1}^N e^{itx_k+isy_k}
\end{equation}

Thus, the sample distance covariance is now given by:
\begin{equation}
\mathcal{V}_N^2(X,Y) = \frac{1}{\pi^2} \int_{\mathbb{R}^2} \frac{|\varphi^N_{XY}(t,s) - \varphi^N_X(t) \varphi^N_Y(s)|^2}{t^2s^2} \, \mathrm{d}t \, \mathrm{d}s
\end{equation}

\citeauthor{Szeetal2007} show how the expression above leads to the final expression in Eq.~\ref{eq:cor}. I follow here the path they drew, but apply it to the case where $Y$ is an angular variable  -- a circular variable ranging between $0$ and $2\pi$. The characteristic function of angular variables is defined only for integer values $m$ \citep{MarJup2000}:
\begin{equation}
\varphi_Y(m) = \int_0^{2\pi} e^{imy} f_Y(y) \mathrm{d}y 
\end{equation}
The empirical characteristic function is now:
\begin{equation}
\label{eq:ecfy}
\varphi_Y^N(m) = \frac{1}{N} \sum_{k=1}^N e^{imy_k}
\end{equation}
The bivariate joint empirical characteristic function for a linear variable $X$ and a circular variable $Y$ is a function of a real variable $t$ and an integer $m$:
\begin{equation}
\label{eq:ecfxy}
\varphi_{XY}^N(t,m) = \frac{1}{N} \sum_{k=1}^N e^{itx_k+imy_k}
\end{equation}

In analogy to the distance covariance for two linear variables, we can now define the linear-circular distance covariance (the phase distance covariance) in the following way:
\begin{equation}
\label{eq:integralandsum}
\mathcal{V}_N^2(X,Y) = \frac{1}{2\pi} \int_\mathbb{R} \sum_{\substack{m=-\infty \\ m \neq 0}}^\infty \frac{|\varphi^N_{XY}(t,m) - \varphi^N_X(t) \varphi^N_Y(m)|^2}{m^2t^2} \, \mathrm{d}t
\end{equation}

The two following observations determine the results of this calculation and also the normalization constant $2\pi$. The first is a special case of Lemma~1 in \citet{Szeetal2007}, which states that for every real $x$:
\begin{equation}
\label{eq:lemma1}
\int_\mathbb{R} \frac{1-\cos(tx)}{t^2} \mathrm{d}t = \pi|x|
\end{equation}
The second is an adaptation to the angular case, which states that for every $y \in [0,2\pi)$:
\begin{equation}
\label{eq:mylemma1}
\sum_{\substack{m=-\infty \\ m \neq 0}}^\infty \frac{1-\cos(my)}{m^2} = \frac{1}{2} y (2\pi-y)
\end{equation}
This last assertion can be proven easily using the properties of the standard Clausen function $\mathrm{Sl}_2(\theta)$ \citep{AbrSte1972}. 

Expansion of the numerator in Eq.~\ref{eq:integralandsum} leads to several expressions of the form of the expressions in Eqs.~\ref{eq:lemma1}~and~\ref{eq:mylemma1}. The origin of the cosine terms is the complex exponent in the definitions of the empirical characteristic functions (Eqs.~\ref{eq:ecfx},~\ref{eq:ecfy}~and~\ref{eq:ecfxy}). Since sine is an odd function, the corresponding sine terms cancel out in the integral and the infinite sum.

After performing the integral and the infinite sum, one is left with algebraic expressions of the kind of Eqs.~2.15--2.18 in \citet{Szeetal2007}. The only difference is that instead of the $Y$-distance $|Y_k-Y_l|_q$, we have the new phase distance $|y_k-y_l|(2\pi-|y_k-y_l|)/2$\,. The last algebraic step which leads to the expressions in Eq.~\ref{eq:distances} and the numerator of Eq.~\ref{eq:cor} is proven in the Appendix of \citet{Szeetal2007}. 

Finally, for convenience, I have converted the angular units in radians to time units by multiplying twice by $P/2\pi$, which does not change the distance correlation value because of the normalization in Equation~\ref{eq:cor}.

\bsp	
\label{lastpage}
	\end{document}